\documentclass[usenatbib]{mn3e} 
\usepackage{amsmath,amssymb}
\usepackage[figuresleft]{rotating}     
\usepackage{tablefootnote}   
\usepackage{graphicx}         
\usepackage{gensymb}        
\usepackage{color}               
\usepackage{natbib}
\usepackage{times}
\usepackage{fixltx2e} 
\usepackage[flushleft]{threeparttable}
\usepackage{hyperref}     
\usepackage{url}
\usepackage{caption}        
\newcommand{\ramses}{{\small RAMSES}}
\newcommand{\ramsestext}{{\small RAMSES} }

\newcommand{\cupidtext}{{\small CUPID}}

\newcommand{\HH}{{\rm H}$_2$ }

\newcommand{\Msol}{{\,{\rm M}}_\odot}

\newcommand{\pc} {{\,\rm pc}} 
 
\newcommand{\cc}{{\,\rm {cm^{-3}}}}

\def\Myr{\,{\rm Myr}}
\def\yr{\,{\rm yr}}
 
\newcommand{\eff}{\epsilon_{\rm ff}}
\newcommand{\effp}{\epsilon_{\rm ff,2D}}
\newcommand{\efflim}{\epsilon_{\rm ff,lim}}
\newcommand{\tff}{t_{\rm ff}}
\newcommand{\tfff}{t_{\rm ff,\,fixed}}
\newcommand{\effsf}{\epsilon_{\rm ff,SF}}
\newcommand{\effmed}{\langle\epsilon_{\rm ff}\rangle_{\rm med}}
\newcommand{\effmedp}{\langle\epsilon_{\rm ff,2D}\rangle_{\rm med}}
\newcommand{\efflee}{\epsilon_{\rm ff,\,L16}}

\newcommand{\tage}{t_{\rm age}}
\newcommand{\tsy}{t_{\rm\star,y}}

\newcommand{\msy}{M_{\rm\star,y}}
\newcommand{\mst}{M_{\rm\star,tot}}
\newcommand{\mtot}{M_{\rm tot}}

\newcommand{\sfepmed}{\langle\epsilon_{\rm ff}\rangle_{\rm med}}
\newcommand{\sfemed}{\langle\epsilon_{\rm ff}\rangle_{\rm med}}

\newcommand{\sigeff}{\sigma_{\log{\eff}}}
\newcommand{\sigeffp}{\sigma_{\log{\effp}}}

\title[Diversity of Star Formation Efficiencies in GMCs] {On the Observed Diversity of Star Formation Efficiencies in Giant Molecular Clouds}\author[Kearn Grisdale
  et al.] {\parbox[t]{\textwidth}{Kearn Grisdale$^{1}$\thanks{kearn.grisdale@physics.ox.ac.uk}, Oscar Agertz$^{2}$, Florent Renaud$^{2}$, Alessandro B. Romeo$^{3}$, Julien Devriendt$^{1}$ and Adrianne Slyz$^{1}$}\vspace*{6pt}\\
  $^1$ Sub-department of Astrophysics, University of Oxford, Keble Road, Oxford OX1 3RH \\
  $^2$ Lund Observatory, Department of Astronomy and Theoretical Physics, Box 43, 221 00 Lund, Sweden\\
  $^3$ Department of Space, Earth and Environment, Chalmers University of Technology, SE-41296 Gothenburg, Sweden\\
  }
\date{\today}
\begin{document}
\maketitle
\graphicspath{ {Figures/} }
\begin{abstract} 
Observations find a median star formation efficiency per free-fall time in Milky Way Giant Molecular Clouds (GMCs) on the order of $\eff\sim 1\%$ with dispersions of $\sim0.5\,{\rm dex}$. The origin of this scatter in $\eff$ is still debated and difficult to reproduce with analytical models. We track the formation, evolution and destruction of GMCs in a hydrodynamical simulation of a Milky Way-like galaxy and by deriving cloud properties in an observationally motivated way, measure the distribution of star formation efficiencies which are in excellent agreement with observations. We find no significant link between $\eff$ and any measured global property of GMCs (e.g. gas mass, velocity dispersion). Instead, a wide range of efficiencies exist in the entire parameter space. From the cloud evolutionary tracks, we find that each cloud follow a \emph{unique} evolutionary path which gives rise to wide diversity in all properties. We argue that it is this diversity in cloud properties, above all else, that results in the dispersion of $\eff$. \\

\end{abstract}



\section{Introduction}
\label{sect:intro}

It is within Giant Molecular Clouds (GMCs) that galaxies form the vast majority of their stars \citep{Myers:1986aa,Shu:1987aa,Scoville:1989aa,McKee:2007aa}. In local spiral and dwarf galaxies, star formation on galactic scales is known to be a slow process \citep[e.g.][]{Bigiel:2008aa}, with gas depletion time scales on the order of billions of years. This inefficiency is also found on scales of individual GMCs \citep{Krumholz:2007aa}, with a median star formation efficiencies per cloud free-fall time $\eff \sim1\%$ \citep[][]{Myers:1986aa,Murray:2011aa,Krumholz:2012aa}. 
	While most observations of $\eff$ in GMCs find very similar median value, different surveys find spreads in $\eff$ of different sizes. Using the median absolute deviation to robustly estimate the standard deviations in $\eff$ the observational data presented in \cite{Evans:2014aa,Heyer:2016aa,Lee:2016aa,Vutisalchavakul:2016aa,Ochsendorf:2017aa} and \cite{Utomo:2018aa} yields values ranging from $\sim0.21$ to $\sim0.83\,{\rm dex}$. 
	 In their recent review \cite{Krumholz:2018aa} compiled data from 13 papers over the last decade (see their Fig.~10) and discuss the impact of the method used to estimate $\eff$ on measurements. The exact size and distribution of the spread in $\eff$ is therefore still a debated topic, particularly as the origins of this diversity is not yet understood.

As the dispersion in $\eff$ is found independently of the method of observation, it is likely physical. Analytical models of star formation in supersonic turbulent flows \citep[e.g.][]{Padoan:2011aa,Hennebelle:2011aa,Federrath:2012aa} have successfully explained the low mean $\eff$ in GMCs and provided insight into how it can scale with global cloud properties, such as density and virial parameter. \cite{Krumholz:2005aa}, for example, postulate that GMCs are turbulent gas structures which are characterised by a log-normal distribution (determined by the Mach number) and only regions within the cloud with density above some threshold are able to form stars. In their model the threshold density is determined by both the virial parameter of the cloud and its Mach number. \cite{Lee:2016aa} \citep[henceforth L16, see also][]{Ochsendorf:2017aa} have been critical of such models for failing to predict a sufficiently large dispersion in $\eff$. Furthermore, observations indicate a decreasing $\eff$ with increasing cloud mass ($M_{\rm GMC}$). \cite{Ochsendorf:2017aa} argued that this is a result of massive clouds having a diffuse, non-star forming outer envelopes. This observed $M_{\rm GMC}$-$\eff$ relationship presents an additional constraint on any model attempting to explain the distribution in $\eff$. However it is worth noting that the $M_{\rm tot}$-$\eff$ relation could be, at least in part, the result of limitations in current observations to detect low mass clouds. 

\cite{Feldmann:2011aa} developed a toy model where the growth of stellar mass in a GMC is determined only by the mass and free-fall time of available gas and an intrinsic efficiency per free-fall time ($\epsilon_{\rm ff,0}$), meanwhile the change in gas mass results from a combination of mass converted into stars, gas removed by feedback and accretion of new gas from the surrounding environment. Using this model they found that a $\sim 2$ dex spread in the measured $\eff$ can be explained by cloud evolution by adopting a \emph{constant} input $\eff$. L16 explored the same effect but found that a fixed input $\eff$ fails to reproduce their observed distribution  of $\eff$, with too few clouds predicted at high ($\gtrsim 10\%$) and low ($\lesssim 0.01\%$) efficiencies at any given time. By assuming that all GMCs evolve in a similar fashion, but observed at a different stage of evolution and by allowing for a time dependent $\eff$, this  problem was mitigated (see their figure 8). While each GMC observed is at a different stage of its life, it is currently unclear if \emph{all} clouds follow the same evolution, and how (or if) the galactic environment plays a role.

A number of authors have studied the dispersion in $\eff$ using simulations of individual GMCs as well as global disc simulations: the former looking for an explanation in the internal properties of clouds, while the latter allows for the impact of environment to be studied.  \cite{Semenov:2016aa} studied the impact of an explicit treatment of small scale gas turbulence, using simulations of entire Milky Way-like galaxies, on the parsec-scale star formation efficiency. From the local properties of the gas, they used the simple parametrisation of \cite{Padoan:2012aa}, calibrated on magneto-hydrodynamical simulations of star formation in supersonic turbulence, to compute $\eff$. They found that their simulations produced values of $0.01\%\lesssim\eff\lesssim10\%$. While this is an encouraging result, we will in this work demonstrate that the mapping between the $\eff$ computed from local properties and what is actually derived from observations is complex and depends on the star formation history of the cloud, not just its instantaneous properties. 

Recently \cite{Grudic:2018aa} carried out 17 magnetohydrodynamic simulations of isolated GMCs of varying mass, radii and feedback models but identical surface density. They found that the spread in $\eff$ seen in observations is similar to the spread in $\eff$ measured throughout a cloud's lifetime. However, as pointed out by the authors, their GMCs have ``fairly artificial'' initial conditions and lack the effects of the larger galactic environment in which GMCs are found. As shown by observations \citep[e.g.][]{Rosolowsky:2003aa,Heyer:2009aa}, GMCs have a large range of properties that can not be captured by 17 overlapping models. Therefore full galactic (disc) simulations, which produce self-consistent GMCs with a range of properties, are needed to better model and investigate the entire evolution of such clouds. 

In this study we will go beyond previous work by using parsec resolution simulations of entire disc galaxies and investigate the emerging GMC star formation efficiencies and how they evolve. This allows us to explore whether such a diversity in $\eff$ and the observed mass--$\eff$ relation is an artefact of observational methods or a physical result. The large number of GMCs found in our simulations allows us to look for correlations between different properties of a GMC and its $\eff$, thus determining if a single property is responsible for the observed scatter in $\eff$. Furthermore, taking advantage of the high temporal resolution of our simulations, we explore how GMCs evolve over their lifetime in a number of different properties and how this contributes to the diversity in $\eff$. Finally, combining our simulated data with analytical models we determine whether a single model is able to explain our simulations or observations.

This paper is organised as the following: in \S\ref{sect:meth} we summarise our simulations and methods for identifying and tracking GMCs, in \S\ref{sect:results} we present the measured values of $\eff$ and how it relates to global GMC properties, in \S\ref{sect:disc} we explore the source of the dispersion in $\eff$ and finally we present our conclusions in \S\ref{sect:con}.

\section{Method}
\label{sect:meth}

\subsection{Simulations}
\label{sect:meth:sims}
We make use of the two Milky Way-like galactic disc simulations in \cite{Grisdale:2016aa}, henceforth G17. The simulations are identical, apart from one being run with stellar feedback (our fiducial simulation) and one without. The simulations account for a dark matter halo, stellar and gaseous disc and a bulge. The initial conditions of both simulations are identical to the AGORA disc initial conditions described in \cite{Kim:2016aa}. They were run using the hydro+$N$-body, Adaptive Mesh Refinement (AMR) code {\ramsestext} \citep{Teyssier:2002aa}. A cell is refined if it reaches a threshold mass of $9300\Msol$ and the minimum allowed cell size is $\Delta x\sim 4.6 \pc$.

The adopted cooling, feedback and star formation models are outlined in G17 and \cite{Grisdale:2018aa}, hence forth G18 \citep[see also][]{Agertz:2013aa,Agertz:2015aa}. Briefly, the feedback model accounts for the injection of momentum, energy, mass loss and enrichment from stellar winds, supernovae (II and Ia) and radiation pressure from young stars. Star formation occurs on a cell-by-cell basis according to the star formation law: 
\begin{equation}
	\dot{\rho}_{\star}= \effsf f_{\rm H_2}\frac{ \rho_{\rm g}}{t_{\rm ff}},
	\label{eq:schmidtH2}	
\end{equation}
where $f_{\rm H_2}$ is the local mass fraction of molecular hydrogen (H$_2$), $\rho_{\rm g}$ is the gas density, $t_{\rm ff}=\sqrt{3\pi/32G\rho_{\rm g}}$ is the local free-fall time and $\effsf$ is the local star formation efficiency per free-fall time of gas in the cell. For all star forming cells $\epsilon_{\rm ff,SF}$ is set to $10\%$ in the simulation with feedback and $1\%$ in the simulation without. As shown in G17, these choices lead to comparable galactic star formation histories in the two simulations. All star particles form with an initial mass of $300\Msol$.

G17 demonstrated that the simulation with stellar feedback give rise to a supersonically turbulent ISM, with a density and velocity structure in close agreement with local spiral galaxies. Furthermore, the resulting GMC population has masses, sizes, velocity dispersions and scaling relations (`Larson's relations') closely matching that of the Milky Way \citep[e.g.][]{Heyer:2009aa}, as shown in G18. This makes our simulations a suitable platform for investigating the evolution and star formation properties of GMCs.

\subsection{Cloud identification and analysis}
\label{sect:meth:id}

We identify GMCs in two separate ways: 1) in projection (2D) using the CLUMPFIND algorithm \citep{Williams:1994aa} as implemented in the clump finding identification and analysis package \cupidtext\footnote{part of the Starlink Project \citep[see][for details]{Starlink2,Starlink}} and 2) in 3D using the on-the-fly clump finding module PHEW \citep[Parallel HiErarchical Watershed,][]{Bleuler:2014aa} built into \ramses. The methods yield similar distributions of GMC properties, albeit with the 3D method giving slightly larger masses and sizes. The adopted cloud finding parameters and the resulting GMC properties are discussed in detail in G18. We adopt both methods when comparing simulated GMC star formation efficiencies in \S\ref{res:dist} but restrict all other analysis in this work to the 3D approach. This allows for higher time resolution, hence allowing for cloud tracking as well as better statistics (due to the number of clouds identified). 
\begin{figure*}
	\begin{center}
		\includegraphics[width=0.98\textwidth]{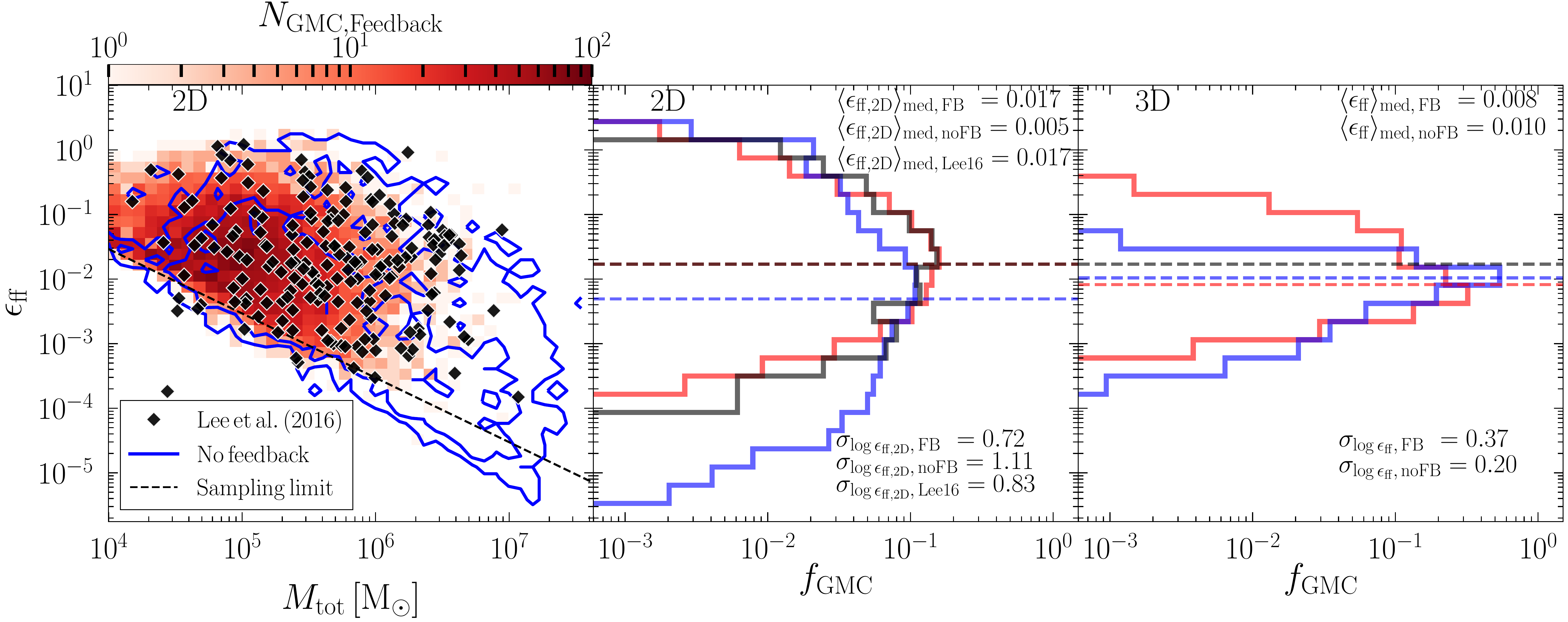}
		\caption[]{Left: $\eff$ as function of total mass ($\mtot$) for clouds identified using the 2D clump finder. GMCs from the simulation with feedback are shown by the red 2D-histogram, while those from the simulation without are given by the blue contours ($\geq1,\,\geq5$ GMCs). The sampling limit of the 2D clump finder is given by the dashed black line (assuming $\tff/\tsy=1$) see text for details.
		 Middle and right: Histograms showing the fraction of GMCs ($f_{\rm GMC}$) with a given $\eff$ for clouds identified using the 2D and 3D clump finders respectively. GMCs from the simulation with feedback are shown in red, while those from the simulation without are shown in blue. 
		 The dashed lines show the median efficiency ($\sfepmed$) for each data set. The values of the $\sfepmed$ and the standard deviation in $\log{\epsilon}$ ($\sigma_{\log\epsilon}$) are given. For comparison with observations, data from Table 3 of L16 is included in the left and middle panels (black points and black histogram respectively) and $\langle\eff\rangle_{\rm med,\,Lee16}$ is shown (dashed black line).  
		} 
		\label{fig:2deff}
	\end{center}
\end{figure*}
Throughout this work, data from the 2D clump finder were obtained from simulation snapshots between $t=150-450\Myr$, separated by  $\Delta t=25\Myr$. For the 3D clump finder, data was obtained at $t=325-380\Myr$, with a temporal spacing of, on average, $\Delta t\sim 25,000$ years (see G17 for more details). Clouds that lie within the central kiloparsec of the galaxy are removed from the analysis. In total, during the period of analysis we identify 8,201 (3,434) GMCs using 2D clump finding in the simulation with(out) feedback and 655,499 (212,056) clouds with the 3D clump finder.

To accurately compare simulations to observations (e.g. L16), we use the estimator 
\begin{equation}
	\eff=\frac{\tff}{\tsy}\frac{\msy}{(M_{\rm GMC} + \msy)}
	\label{eq:sfeff}
\end{equation}
for each GMC, where $\msy$ is the mass of stars with the GMC which has an age less than $\tsy$,  $M_{\rm GMC}$ is the (molecular) gas mass of the GMC, $\tff=\sqrt{3\pi/32G\rho_{\rm GMC}}$ is the mean free fall time across the GMC and $\rho_{\rm GMC}$ is $M_{\rm GMC}$ divided by the GMC's volume (i.e. its mean density). Stellar masses are calculated by considering only stars that overlap with gas belonging to a GMC\footnote{Other methods for matching stars to GMCs were explored and found to have little impact on the results presented in this work.}, either in projection (2D method) or in 3D. 

To allow for a comparison to the stellar clusters detected by free--free emission (e.g. L16), we adopt $\tsy=4\Myr$. We emphasise that $\eff$ should not be confused with $\epsilon_{\rm ff,SF}$ in Eq.~\ref{eq:schmidtH2}. The former is the efficiency per free fall time averaged over the whole GMC ($\sim10-70\pc$), while the latter is the efficiency at which an \emph{individual computational cell} converts gas into stars ($\sim4.6\pc$). 

\subsection{Cloud tracking}
\label{sect:meth:track}
All GMC quantities are followed over time by employing the cloud tracking algorithm described in  \cite{Tasker:2009aa}. Briefly, for clouds found using the 3D\footnote{We only apply the tracking routine to clouds identified with the 3D clump finding method, as the time resolution in the 2D method is insufficient for accurate tracking.}, we use the position ($\mathbf{x}$) and velocities ($\mathbf{v}$), of each GMC at a time $t_{0}$, we adopt a linear approximation, where the change in position vector over one tracking step is $\Delta \mathbf{x}=\mathbf{v}\Delta t$, to predict where the cloud should be at the next cloud finder output, $t_{1}=t_0+\Delta t$. Next, separations ($\mathcal{S}$) between a cloud's predicted position at $t=t_{1}$ and the positions of all clouds found at this time are calculated. Cloud identities are matched for clouds with the smallest $\mathcal{S}$ and satisfying either $\mathcal{S} \leq2R_{{\rm GMC},t_{0}}$ or $\mathcal{S} \leq\langle R_{{\rm GMC},t_{1}}\rangle$, where $R_{{\rm GMC},t_{0}}$ is the radius of the cloud at $t_{0}$ and  $\langle R_{{\rm GMC},t_{1}}\rangle$ is the mean cloud radius at $t_{1}$. In the case of multiple clouds from $t_{0}$ being linked to the same cloud at $t_{1}$, the cloud at $t_{1}$ inherits the identity of the most massive cloud from $t_{0}$, while the other progenitor cloud(s) are considered to have merged and are not tracked further.

To ensure that only clouds with a complete life-cycle are considered we exclude those formed in (or before) the first snapshot of our analysis, or those destroyed after the last. 

Furthermore, it is important to note that because we are detecting clouds based on a fixed density threshold ($100\cc$, see G18) it is possible for a cloud to drop below the detection limit but remain a coherent structure and then, at later times, pass back above the threshold. In such situations, our methods would register the cloud as having been destroyed and a new cloud forming. A ``new'' cloud of this type may be detected with a significant young stellar mass ($\msy$), hence giving the appearance of beginning its life with a high $\eff$. To mitigate this, we only considered GMCs with an initial $\msy\leq1500\Msol$, equivalent to five (or less) star particles. An alternative method, which would better reflect the complex cycle of gas ending up in GMCs, would be to employ tracer particles to track the gas of each GMC \citep[e.g.][]{Semenov:2016aa}, or to identify stellar clusters and their associated molecular gas. We leave this for a future investigation. 

Finally, clouds with lifetimes shorter than a million years are also discarded. After the tracking is complete, and the above criteria are applied, 1879 unique GMCs evolutionary tracks remain, which we focus on in \S\ref{res:evol}.

\section{Results}
\label{sect:results}
\subsection{Distribution of star formation efficiencies}
\label{res:dist}

\subsubsection{The GMC Mass - Star Formation Efficiency Relation}
\label{res:masssffe}
We begin our analysis by calculating $\effp$ (Eq.~\ref{eq:sfeff}) for the clouds identified in projection, as outlined in \S\ref{sect:meth:track}. In the left hand panel of Fig.~\ref{fig:2deff} we show how $\effp$ varies with total cloud mass ($\mtot=M_{\rm GMC}+\msy$, as defined in L16) in our fiducial simulation (i.e. including feedback) and compare these to the observational data in L16. The simulated GMC population has a wide range of $\effp$ values which agree well with observations, see Fig.\ref{fig:2deff}. 

Both simulated and observed $\effp$ decrease with increasing $\mtot$ \citep[see also][]{Ochsendorf:2017aa}, raising the question as to whether this is due to a physical process, or a result of the cloud identification method. In our simulation $\msy=N\cdot M_{\star}$, where $M_{\star}=300\Msol$\footnote{Particles in the feedback simulation lose mass, however during the first $4\Myr$ of their evolution they only lose a maximum of $10\%$ of their initial mass.}
 is the mass resolution of the star particles and $N$ is the number of star particles in the GMC. This defines a lower limit of the estimated $\effp$ in our simulation,
\begin{equation}
	\efflim  = \frac{\tff}{\tsy}\frac{1}{(1+ M_{\rm GMC}/300\Msol)},
	\label{eq:quanta}
\end{equation}

below which our simulation cannot sample star formation. This `sampling limit' is shown in Fig.~\ref{fig:2deff} for $\tff=\tsy$, illustrating how this introduces a bias in how $\effp$ relates to $\mtot$. Star clusters identified via free--free emission, as done by L16, have a similar bias; such identification requires the presence of UV-emitting massive stars to ionise the surrounding ISM, hence setting a lower limit to the detectable star formation efficiency for all cloud masses, which scales in a similar fashion as Eq.~\ref{eq:quanta} \citep[see also][]{Murray:2011aa,Grudic:2018aa}. This is likely a contributing factor as to why the simulation and observations agree on the low $\effp$ end of the distribution. Indeed, with the exception of a single GMC,\footnote{The exception, found with $\eff=1.8\times10^{-4}$ and $\mtot=2.8\times10^{4}\Msol$, only has  $16\Msol$ in stellar mass. }  the observed GMCs shown in Fig.~\ref{fig:2deff} (from L16) have $\msy\gtrsim100\Msol$, close to the mass resolution of the simulation.

\subsubsection{Dispersion of Star Formation Efficiencies}
\label{res:disp}
To quantify the spread of $\effp$ we show the normalised distribution, the median value of $\effp$ ($\effmedp$) and the standard deviation of $\log{\effp}$, $\sigeffp$\footnote {The standard deviation ($\sigeffp$) is robustly estimated via the median absolute deviation (MAD): $\sigeffp= 1/0.6745$ MAD \citep{Muller:2000aa,Romeo:2016aa}.}, for both the simulation and L16's observations in the middle panel of Fig.~\ref{fig:2deff}. 
\begin{figure}
	\begin{center}
		\includegraphics[width=0.4\textwidth]{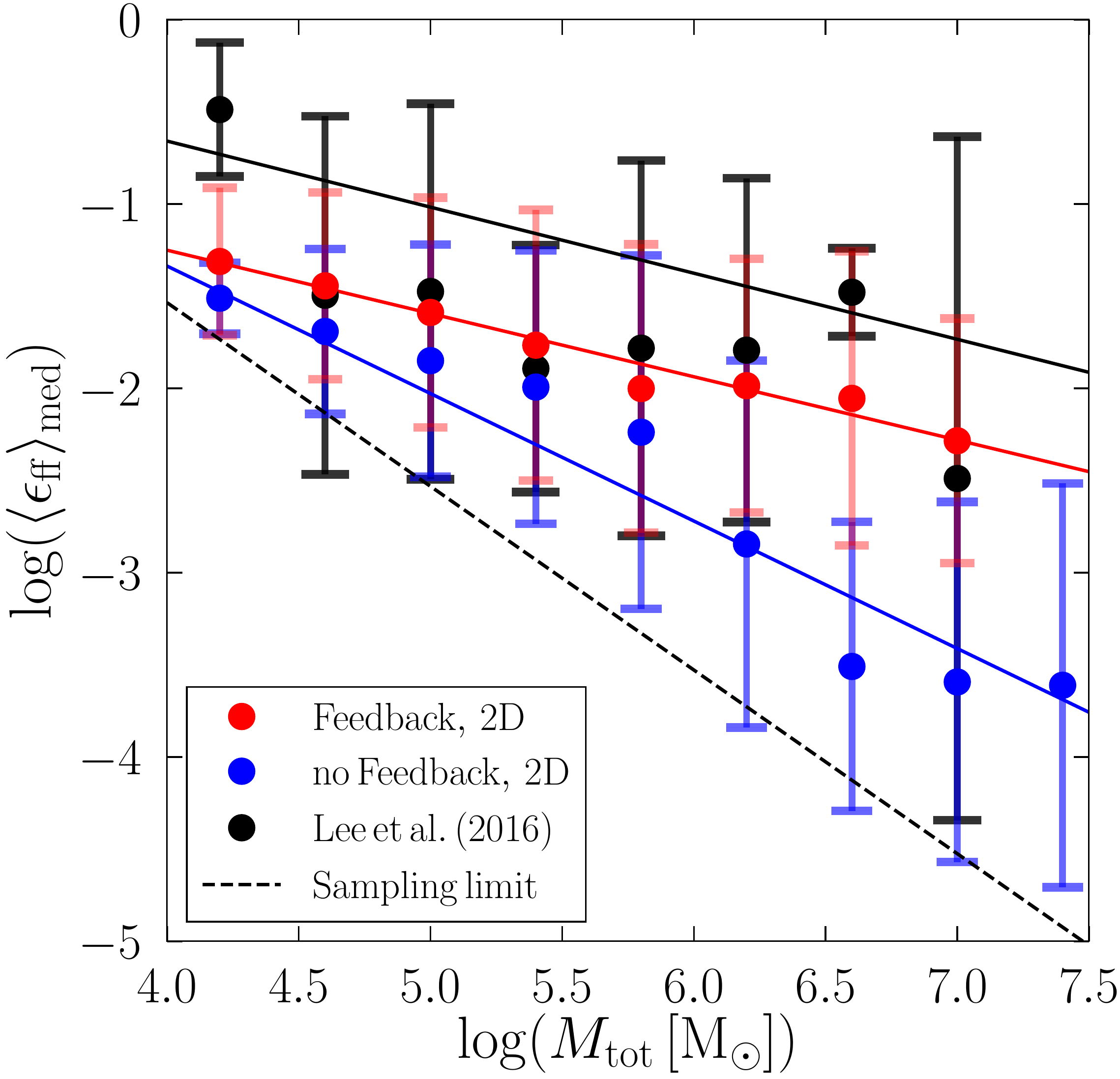}
		\caption[]{$\effmed$ as a functions of $\mtot$. Shown are measurements for GMCs identified using the 2D clumpfinder in the simulation with feedback, without feedback and from L16, shown using red, blue and black points respectively. The error bars show $\sigeff$ for each mass bin. The solid red, blue and black lines show a $\chi^{2}$ least squares fit to their corresponding data set. We note the fits do not account of the sampling limit shown by the dashed-black line (see text).
		}
		\label{fig:effmass}
	\end{center}
\end{figure}
\begin{figure*}
	\begin{center}
		\includegraphics[width=0.90\textwidth]{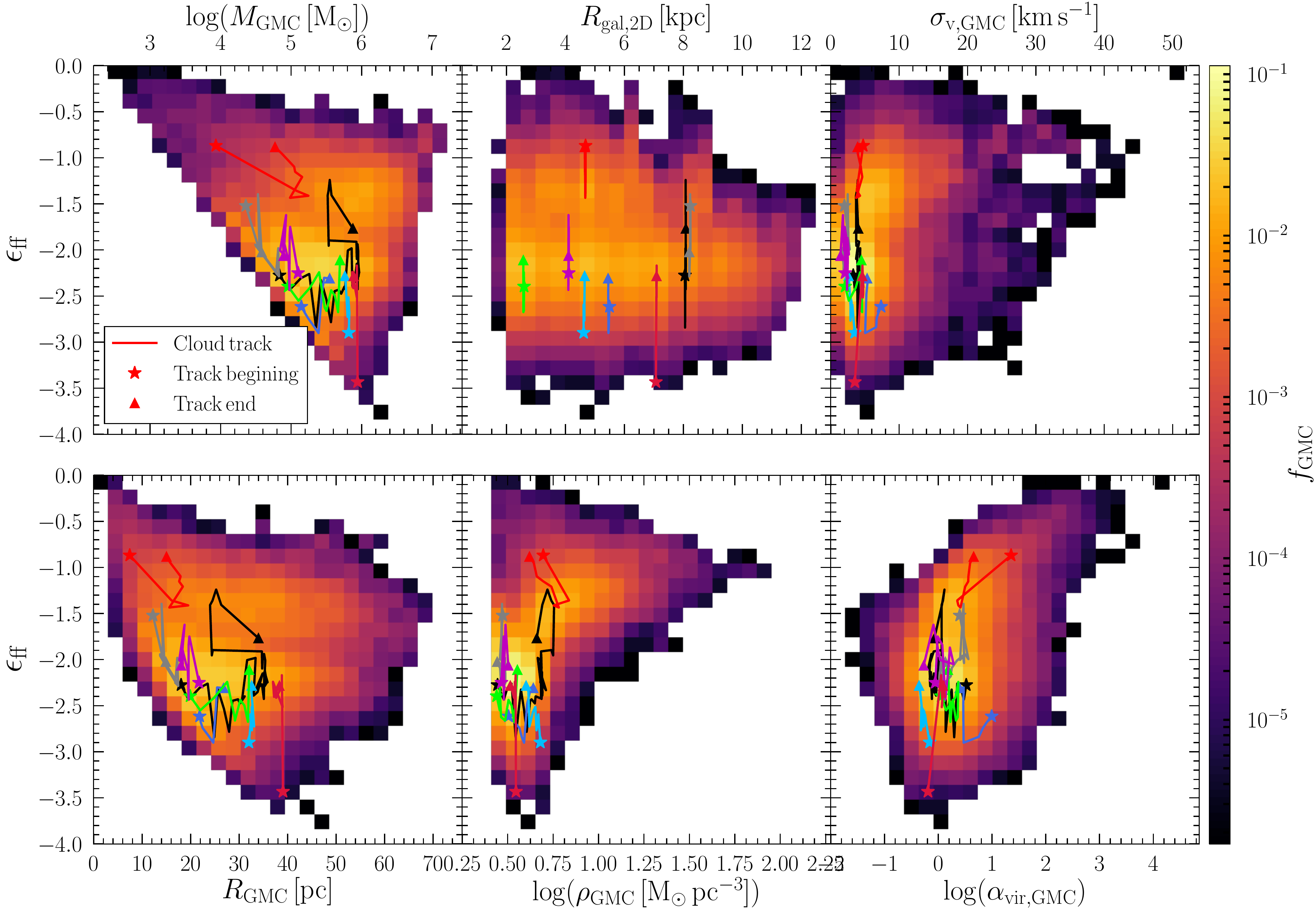}
		\caption[]{2D  Histograms showing how the distribution of $\eff$ is correlated with $M_{\rm GMC} $, $R_{\rm gal,2D}$, $\sigma_{\rm v,GMC}$, $R_{\rm GMC}$, $\rho_{\rm GMC}$ and $\alpha_{\rm vir,GMC}$ of the GMCs identified with the 3D clump finder in the simulation with feedback. All histograms use the same colour scale (shown on the right), which is normalised to the total number of GMCs included in the data. Additionally, each panel shows the evolution tracks of eight randomly selected GMCs.
		}
		\label{fig:effprop}
	\end{center}
\end{figure*}
The shape of the $\effp$ distribution in our simulation is in excellent agreement with observations. We find a median star formation efficiency per free-fall time of $\sim 2~\%$, also matching observations. The simulated $\sigeffp$ is not as large a value as L16's data ($0.72\,{\rm dex}$ compared to $0.83\,{\rm dex}$) but is within the range of values reported in the literature (see \S\ref{sect:intro}). This is likely due to the simulated population of GMCs being better sampled at low masses compared to observation (see also G18), where the spread in $\effp$ is smaller, as well the observed population having a couple of extremely inefficiently star forming clouds ($\eff\sim10^{-4}$). From Fig.~\ref{fig:effmass}, which shows how $\effmedp$ and $\sigeffp$ vary with $\mtot$, we find that $\effmedp$ in the simulation is compatible with observations at almost all cloud masses: low mass clouds ($\sim$ few $\times 10^4 \Msol$) reaching almost $\eff\sim 10\%$, whereas clouds in excess of $10^6\Msol$ have $\eff\lesssim 1\%$, on average. This result can be summarised as $\eff\propto M_{\rm tot}^{\beta}$ with $\beta\sim-0.34$ and $-0.36$ for our simulation and L16's observations respectively. Finally, a qualitative agreement is found for $\sigeffp-\mtot$, with an increasing scatter with increasing cloud mass.

It is important to note that Fig.~\ref{fig:effmass} and the values of $\beta$ given above do not account for the sampling limit (see \S\ref{res:masssffe}). As shown by the L16's data in Fig.~\ref{fig:2deff}, clouds below this sampling limit do exist and therefore need to be accounted for when calculating true value of $\beta$. For example, if we ignore clouds in our feedback simulation that sit on or close to the sampling limit we find $\beta$ decreases, i.e. the relationship become steeper. The steeping in due to the fact that sampling limit preferentially impacts lower mass ($<10^{5}\Msol$ ) clouds, indeed $\effmedp$ at these masses is only $\sim 1\sigeffp$ from the sampling limit. Given that $\beta$ will be biased in both simulation and observation, determining the impact of the sampling limit is of significant importance. One way to assess the how the sampling limit biases measurements of $\beta$ would be to compare data with a simple Bayesian model. Such a comparison is beyond the scope of this study and we leave it for future work.

We calculate the efficiency per free-fall time ($\eff$) for GMCs identified using the 3D clump finder and repeat the above analysis, see right most panel of Fig.~\ref{fig:2deff}. We find similar median efficiencies as before ($\effmed\sim 1\%$), but the spread in $\eff$, which still covers several orders of magnitude, is smaller: $\sigeff=0.37$. 

In summary, the \emph{measured} star formation efficiency per free-fall time in a cloud tells us little, if anything at all, about the \emph{input} efficiency on smaller scales, i.e. $\sfemed\ne\effsf$. The former depends not only on the turbulent density substructure (as discussed below) and global properties of the cloud but also on its evolution and therefore  the rate of star formation, that results in the young stellar population observed at any instance. 

\subsubsection{Role of Feedback on $\eff$}
Fig.~\ref{fig:2deff} and \ref{fig:effmass} include analysis from the simulation without feedback. We find very similar trends as before, i.e. $\effp\propto M_{\rm tot}^{\beta}$ with $\beta\sim-0.69$ and a smaller scatter in 3D compared to 2D (i.e. $\sigeffp=1.1\rightarrow\sigeff=0.2$). As with simulation with feedback we emphasise that the value of $\beta$ given above does not account for the sampling limit and therefore may not accurately represent the correlation between $\effp$ and $M_{\rm tot}$. Given that the simulation without feedback has a propensity to produce unphysical (see G18), massive ($M_{\rm GMC}>10^{7}\Msol$), low $\effp(\lesssim10^{-5})$ clouds, we expect that clouds will be below the sampling limit at all masses, but particularly at high mass (see left panel of Fig.~\ref{fig:2deff}) which makes predicting the impact of the sampling limit on $\beta$ much more difficult in this case.

The primary difference between the GMCs from the simulation with feedback and the simulation without is the latter has a population of long lived massive ($\gtrsim 10^7\Msol$) and very inefficiently star forming  ($\eff\lesssim10^{-3}$) clouds. Interestingly, we find that  in 3D the simulation with feedback yields GMCs able to reach higher $\eff$ values than clouds in the simulation without, which is a result of stellar feedback removing gas from the clouds. From these results we conclude that, while stellar feedback plays a role in determining the shape of the distribution of measured $\effp$ and $\eff$, it is not the source of the dispersion in either.

Given the reasonable match between GMCs in the simulation without feedback and the observations, it might seem just as reasonable to use this simulation as the simulation with feedback in further analysis. However, as shown in both G17 and G18, this simulation fails to produce a realistic neutral ISM and distribution of GMC properties, therefore we focus all further analysis on the simulation with feedback. 

Having established that star formation efficiencies in the simulated GMC population closely match observations, we next aim to quantify why this is the case.
\begin{figure}
	\begin{center}
		\includegraphics[width=0.4\textwidth]{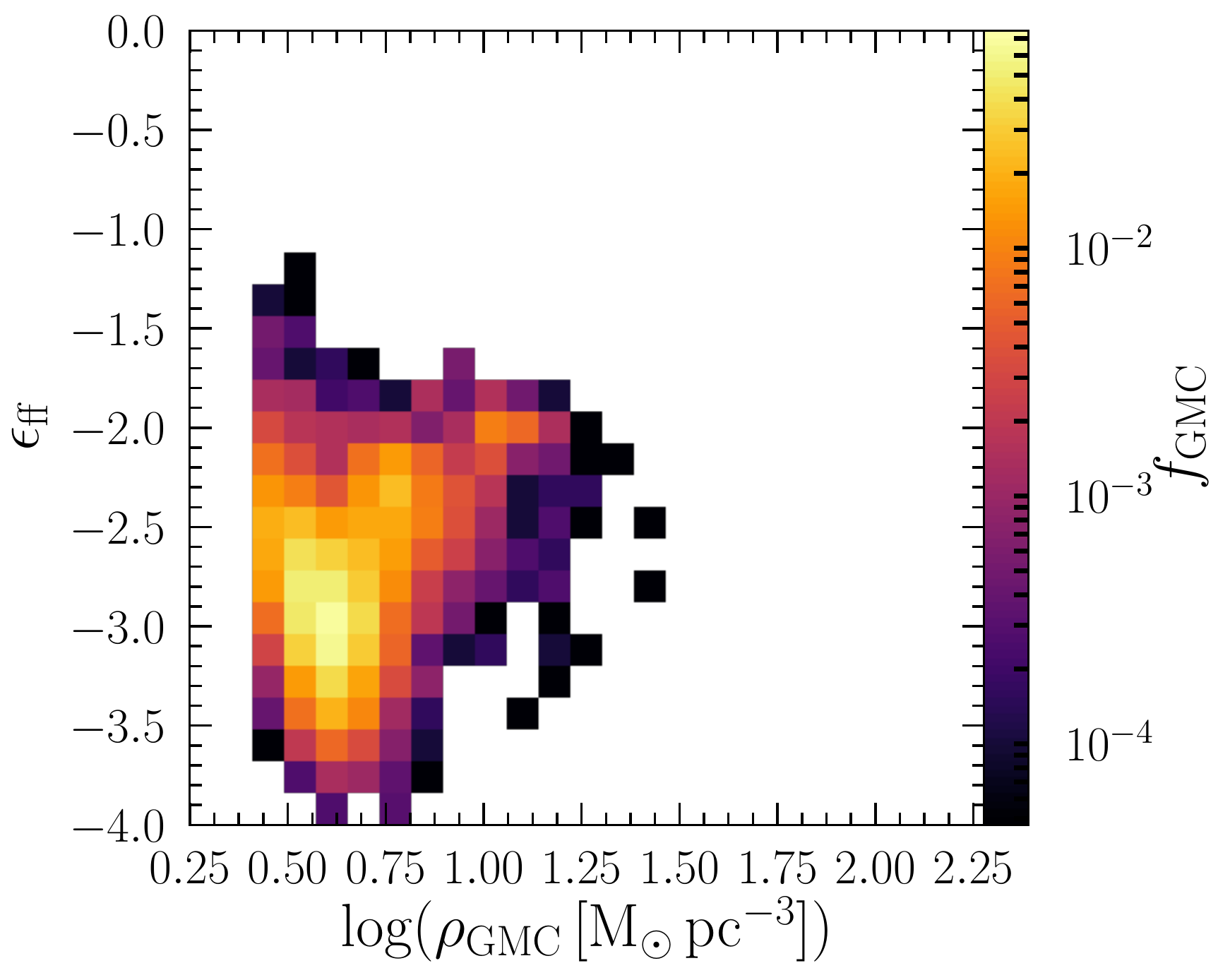}
		\caption[]{2D  Histograms showing how the distribution of $\eff$ is correlated with $\rho_{\rm GMC}$ for the GMCs identified with the 3D clump finder in the simulation without feedback. 
		}
		\label{fig:effpropnofb}
	\end{center}
\end{figure}
\subsubsection{Role of Cloud Properties on $\eff$}
\label{res:evp}
From this point forward we focus our analysis to clouds identified using the 3D clump finder. Fig.~\ref{fig:effprop} shows the relation between $\eff$ of GMCs and gas mass ($M_{\rm GMC}$), galactocentric radius ($R_{\rm gal, 2D}$), velocity dispersion ($\sigma_{\rm v,GMC}$), size ($R_{\rm GMC}$), gas density ($\rho_{\rm GMC}$) and virial parameter ($\alpha_{\rm vir,GMC}$) in the simulation with feedback (we refer the reader to G18 for discussion on how these quantities are calculated). 
\begin{figure*}
	\begin{center}
		\includegraphics[width=1.0\textwidth]{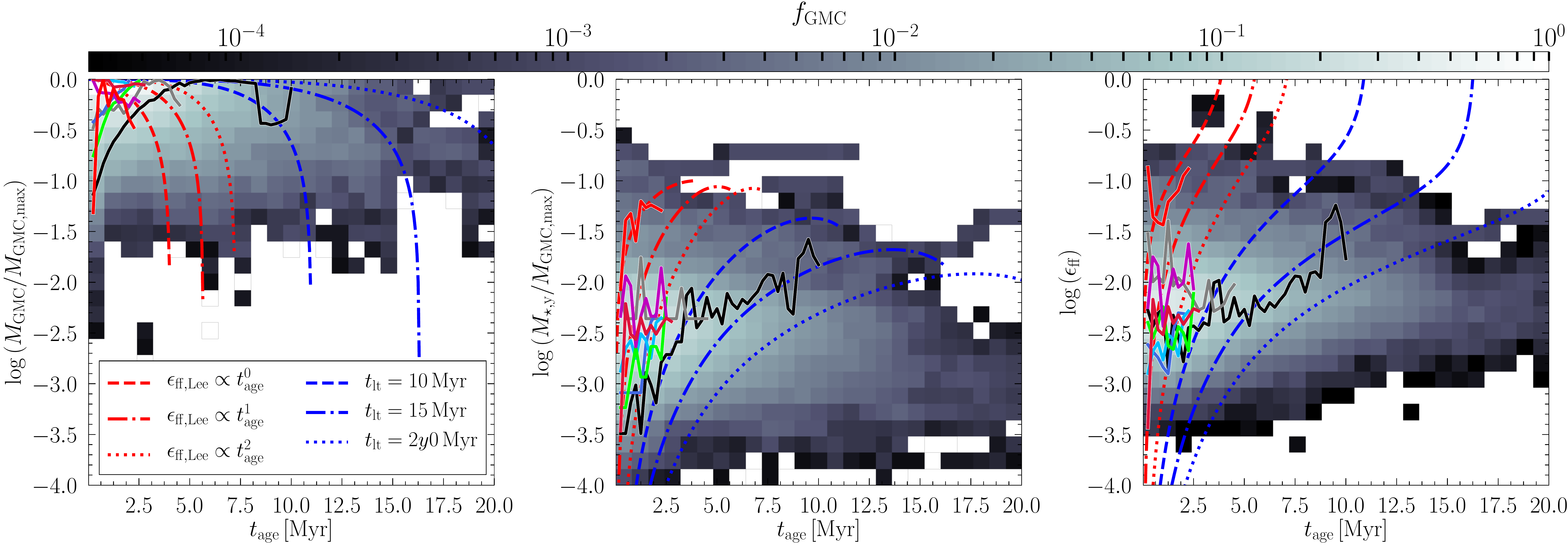}
		\caption[]{Distribution of $M_{\rm GMC},\,\msy,$ and $\eff$ as function of GMC age ($\tage$). We normalise the y-axes by the maximum recored gas mass of GMC during its lifetime ($M_{\rm GMC,max}$). The three red lines show the predicted evolution given by the L16 model (see \S\ref{disc:con}) for $\efflee\propto \tage^{\delta}$ where $\delta=0,\,1$ or $2$, while the blue lines show evolution of clouds with different lifetimes ($10,\,15$ and $20\Myr$) and $\delta=2$. Also shown are the evolution tracks for the same eight GMCs shown in Fig.~\ref{fig:effprop}.
				}
		\label{fig:eff2Dhisttracks}
	\end{center}
\end{figure*}
No strong correlation is found between any cloud property and $\eff$; clouds with similar physical parameters show a great diversity of $\eff$. Not surprisingly, there is a trend for dense GMCs to have higher $\eff$ and likewise, as $t_{\rm ff}\propto \rho^{-0.5}$, a trend for clouds with short free-fall times to be efficient at forming stars. We find a signature of clustering in the different $\eff$-spaces, e.g. in the $\eff-M_{\rm GMC}$ space, clouds cluster around $[\eff,M_{\rm GMC}]\sim [-1.25,10^{6}\Msol]$ and $\sim [-2.3,10^{5.4}\Msol]$, suggesting there are some preferential values that clouds are drawn to. 

Given that $\eff$ in the former of these two regions is centred on $\eff\sim5\%$ and therefore within a factor of 2 from $\effsf$, this region could be an artefact of the star formation model employed in the simulations. To test this we reproduce Fig.~\ref{fig:effprop} using clouds identified in the simulation without feedback (Fig.~\ref{fig:effpropnofb} shows the $\eff$-$\rho_{\rm GMC}$ panel). While the vast majority ($\sim70\%$) of clouds are highly inefficient at forming stars at higher density, i.e. $\rho_{\rm GMC}\gtrsim10\Msol\pc^{-3}$, there is a tendency  for $\eff\rightarrow\effsf$. By directly comparing the $\eff$-$\rho_{\rm GMC}$ panel in Fig.~\ref{fig:effprop} to Fig.~\ref{fig:effpropnofb} we find that general shape of the two distributions is very similar, but that latter is missing most of the $\eff>10^{-1.5}$ clouds. We therefore conclude that the high $\rho_{\rm GMC}$-high $\eff$ region is at least partially a result of the star formation prescription employed in our simulations. In future work we explore in detail, to what the degree the star formation prescription drives clouds into this region and if other factors play a role.

Next we explore how cloud properties evolve and the role this plays in establishing the wide range of observed star formation efficiencies.

\subsection{Cloud evolution}
\label{res:evol}
\subsubsection{Individual clouds}

Fig.~\ref{fig:effprop} includes evolutionary tracks of eight randomly selected GMCs, with their positions shown every $0.25\Myr$. Each cloud has a unique path through the seven parameter spaces explored. For example, most clouds tend to become more gravitationally bound over their lifetime ($\alpha_{\rm vir,GMC}$ decreases) while the crimson cloud becomes less bound ($\alpha_{\rm vir,GMC}$ increases). 

GMCs are not confined to a single area of parameter space but can move from one region to another and in a variety of ways. Comparing the black and grey clouds with either of the blue clouds shows that some clouds explore only a small fraction of a given parameter space while others might explore a large portion. The one possible exception to this is a cloud's progression in $\eff$--$R_{\rm gal,2D}$ space, where we see that clouds are ``born'' and ``die'' at approximately the same galactic radius ($R_{\rm gal,2D}$), with very little variations over the clouds lifetime. This is due to the short cloud evolution timescale ($\sim10\Myr$) compared to the galactic dynamical timescale ($\sim100-200\yr$). Cloud evolution in the $\eff$--$\sigma_{\rm v,GMC}$ space is similar to $\eff$--$R_{\rm gal,2D}$, i.e. evolution in $\eff$ occurs while $\sigma_{\rm v,GMC}$ remains largely unchanged.

A visual inspection of the evolution of these eight clouds 
(see additional material\footnote{Additional material can be found at \url{https://www.youtube.com/playlist?list=PLO1VCxfFwCjDefBvKCF7-CptcZLn-gGMT}}) 
reveals that the environment of a GMC is as important as its internal processes. For example, the red cloud is situated in a particularly dense spiral arm which feeds the cloud with gas, allowing $M_{\rm GMC}$ to increase by almost an order of magnitude during the first $0.5\Myr$ of its life.  In contrast the grey cloud forms in a much lower density environment, which is quickly disrupted by shear from galactic rotation. Furthermore only one of the eight clouds (dark blue) is clearly destroyed by supernovae, with the other seven instead appearing to be destroyed by shear, demonstrating that environment plays a role in the evolution of a cloud throughout its life. 

\subsubsection{General Trends in Evolution}

To aid in teasing out general trends in cloud evolution, we create 2D histograms of $M_{\rm GMC},\, \msy$ and $\eff$ as functions of GMC age ($\tage$) for \emph{all} 1,879 evolutionary tracks (see \S\ref{sect:meth:track}) in Fig.~\ref{fig:eff2Dhisttracks}. There is a wide range of different evolutionary paths taken by GMCs and that most clouds only live for $3$--$4\Myr$.  Furthermore, this demonstrates that the eight randomly selected GMCs overlaid in both Fig.~\ref{fig:effprop} and Fig.~\ref{fig:eff2Dhisttracks} are not the only GMCs with unique evolutionary paths. 

The gas mass of a GMC when it is first detected \emph{tends} to be the maximum gas mass ($M_{\rm GMC,max}$) that the cloud reaches. However, some GMCs reach their $M_{\rm GMC,max}$ at $\tage\ne0$ and have therefore gained mass through cloud-cloud collisions and accretion. Despite this, the general trend is for GMCs to lose mass as they evolve. 

Interestingly, \cite{Kawamura:2009aa} infer from observations that $M_{\rm GMC}(\tage=0)\ne M_{\rm GMC,max}$ and that over a period of $\sim30\Myr$ that $M_{\rm GMC}$ increases by a up to a factor of three, even after star formation begins (see their \S4.2 and Table 4). This discrepancy could be due to any number of factors such as: how we account for cloud-cloud interactions, clouds dropping below our detection threshold (see \S\ref{sect:meth:track}), differences in how we define and detect clouds or inaccuracies in the model used to infer evolution of GMCs from observations. We leave further exploration of this difference to future work.

An obvious assumption would be that the lost gas mass is converted into stars. This is at least partially true, as as the young stellar mass ($\msy$) tends to increase during the first few million years of a clouds life. At $\tage\gtrsim4\Myr$, $\msy$ decreases with increasing age, yet clouds continue to lose significant fractions of their gas mass. Therefore,  the lost gas mass is removed from the GMC by other means (e.g. feedback and shear). 

In general we find that $\eff$ tends to increase during the first $4$--$6\Myr$ of a GMC's life after this point we see that $\eff$ tends to either plateau or decrease. This corresponds to the age at which star particles will experience their first supernova event and therefore is a strong indicator of feedback limiting $\eff$. 

To summarise, at any given $\tage$ it is possible for clouds to have a large variety in $\eff$, $M_{\rm GMC}$ and $\msy$ and the exact value of given property at a given age is unique to each cloud.
\begin{figure*}
	\begin{center}
		\includegraphics[width=1.0\textwidth]{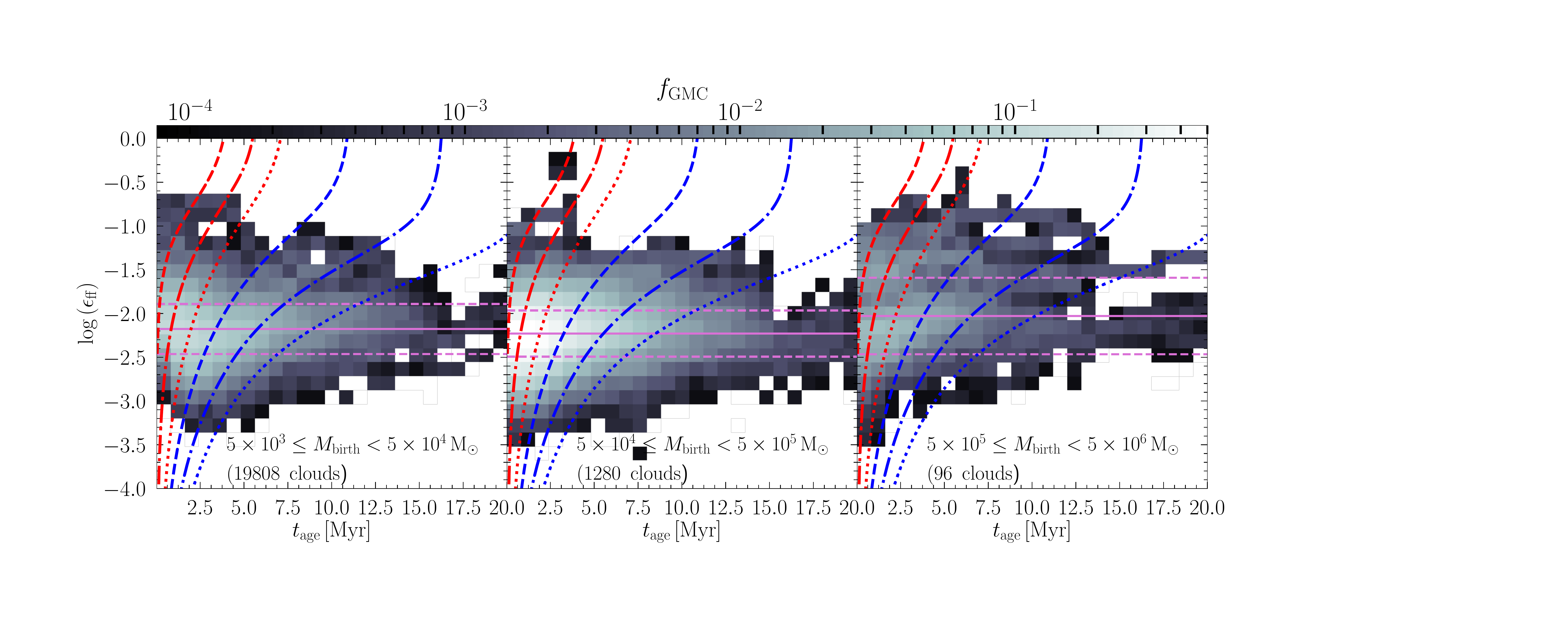}
		\caption[]{2D histogram of $\eff$ as function of GMC age ($\tage$) for GMCs with different ``birth'' masses ($M_{\rm birth}$). The three red lines show the predicted evolution of GMC as given by the L16 model (see \S\ref{disc:con}) for $\efflee\propto \tage^{\delta}$ where $\delta=0,\,1$ or $2$,  while the blue lines show evolution of clouds with different lifetimes ($t_{\rm lt}=10,\,15$ and $20\Myr$) and $\delta=2$, as in Fig.~\ref{fig:eff2Dhisttracks} The purple solid and dashed line in each pane shows $\effmed$ and $\effmed\pm\sigeff$ for each mass range (ignoring GMC age).
		}
		\label{fig:masscuts}
	\end{center}
\end{figure*}
\section{Discussion}
\label{sect:disc}
	 
\subsection{Cloud Conformity or Diversity?}
\label{disc:con}

As shown by the observational data present in L16 and included in Fig.~\ref{fig:2deff} and discussed in \S\ref{sect:intro}, there is a significant dispersion in the values of $\eff$ for GMCs in the Milky Way. \cite{Feldmann:2011aa} put forward a model that is able to produce a wide spread in measured $\eff$ values ($1\lesssim\eff\lesssim100\%$, see their Fig.~2) by adopting a fixed universal efficiency per free-fall time $\epsilon_{\rm ff,0}$ (analogous to $\epsilon_{\rm ff,SF}$ used in our simulations) and allowing the GMC to evolve with time. L16 combined \cite{Feldmann:2011aa}'s model with the star formation prescription given in \cite{Krumholz:2005aa} to allow for a time-dependant $\epsilon_{\rm ff,0}$ which results in a pair of coupled ordinary differential equations: \\
\begin{equation}
	\frac{dM_{\rm GMC}}{dt}= -\epsilon_{\rm ff,0}\left(\frac{\tage}{\tfff}\right)^{\delta}\frac{M_{\rm GMC}(\tage)}{\tfff} -\alpha \mst(\tage)+ \gamma,
	\label{eq:lee1}
\end{equation}
and
\begin{equation}
	\frac{d\mst}{dt}= \epsilon_{\rm ff,0} \left(\frac{\tage}{\tfff}\right)^{\delta}\frac{M_{\rm GMC}(\tage)}{\tfff},
	\label{eq:lee2}	
\end{equation}
where $\gamma$ is the rate of gas accretion on to the GMC and $\alpha$ is a parametrisation of the disruption of GMCs due to feedback. Having solved the above equations the young stellar mass   
\begin{equation}
	\msy(\tage) = \mst(\tage) - \mst(\tage-\tsy),
	\label{eq:lee3}
\end{equation}
and the model  equivalent of $\eff$ 
\begin{equation}
	\efflee(\tage)= \frac{\tfff}{\tsy}\frac{\msy(\tage)}{M_{\rm GMC}(\tage)+\mst(\tage)},
	\label{eq:lee4}
\end{equation}
can be calculated at each at every $\tage$. 
In this model, the evolution of $M_{\rm GMC}/M_{\rm GMC, max},\, \msy/M_{\rm GMC, max}$ and $\efflee$ is the same for  \emph{all} clouds, for a given choice of $\delta$, $\epsilon_{\rm ff,0}$ and fixed free-fall time ($\tfff$).  In essence, this model requires conformity in the evolution of all GMCs and the dispersion in $\eff$ is produced by observing a population of GMCs, with each cloud at a different stage in its evolution. 

We adopt the values of $\alpha,\,\gamma$ and $\tfff$ given in L16 ($3.5,\,0$ and $6.7$  respectively) and reproduce their model for $\delta=0,1$ and $2$. L16 employed $\epsilon_{\rm ff,0}=0.014$ to ensure that all models produce clouds with lifetimes of $\sim20\Myr$ and argued that $\epsilon_{\rm ff,0}\propto\tage^{\delta}$ with $\delta=2$ was required to match observations.  We adopt several different values for $\epsilon_{\rm ff,0}$. Firstly, we use $0.27,\,0.52$ and $0.47$ for the three values of $\delta$ respectively as these values ensure that all models convert $10\%$ of their gas mass to stars by the time the cloud is destroyed ($M_{\rm GMC}/M_{\rm GMC, max}\leq0.01$).  This ensures that all models have the same initial and final conditions (i.e. enforces conformity between models). Thus allowing us to determine if the diversity in our simulated GMCs can be explained by conformity to a single evolutionary path. The second set of values, $0.091,\,0.019$ and $0.006$, produce cloud lifetimes of $10,\,15$ and $20\Myr$ respectively for $\delta=2$ (i.e. assuming that GMCs will have different evolutionary paths). This set of models will allow for a determination on whether simply allowing for different GMC lifetimes is enough to explain the diversity in clouds efficiencies. 

Testing this model against the evolutionary tracks of all GMCs in our simulation, i.e the red and blue lines in Fig.~\ref{fig:eff2Dhisttracks}, shows that due to the large spread in our data, the model (independent of $\delta$ and $\epsilon_{\rm ff,0}$) overlaps with our simulated GMCs in each parameter space. However, that to reproduce the diversity seen in the simulated GMCs would require a significant number of models, each with a different values for $\delta$ and $\epsilon_{\rm ff,0}$: therefore a model which produces diversity is required.  

In their recent work, \cite{Grudic:2018aa} carried out isolated GMC simulations for three different mass clouds ($2\times10^{4,\,5,\,6}\Msol$). They found that by observing a population of clouds, all with the same mass, at random points during their lifetime they could reasonably reproduce L16's observed distribution in $\eff$. Furthermore, this was found independently of cloud mass (see their Fig.~4). Their simulations and conclusion support the models presented in L16: i.e. all clouds follow (nearly) identical evolutions and the spread in $\eff$ is a result of observing a population of different aged clouds.

Fig.~\ref{fig:masscuts} shows how $\eff$ evolves for clouds with gas mass at ``birth''\footnote{Defined as the first time a cloud is detected by the clump finder} ($M_{\rm birth}$).
The clouds with the largest $M_{\rm birth}$ tend to reach higher $\eff$ and have a (marginally) higher $\effmed$. The most noteworthy result from Fig.~\ref{fig:masscuts} is that birth mass plays only a small role in determining the initial value of $\eff$ as shown by $\eff(\tage\leq0.8\Myr)$ having at least a 2 dex spread in each mass bin. Furthermore, while the dispersion in $\eff$ decrease as clouds age, it never reaches zero. This implies that knowing the birth mass of a GMC is not enough to predict the efficiency at which it converts gas into stars, and thus its star formation history.
Carrying out a similar experiment using $R_{\rm birth}$ (not shown), yields nearly identical results: clouds are born with a range of different $\eff$ independent of their initial size. 

Unlike the model presented in L16 and the isolated GMCs simulations of \cite{Grudic:2018aa}, the GMCs in this work are simulated in a (realistic\footnote{As shown by the analysis of the simulation in G17 and G18.}) 
galactic environment which can heavily influence their evolution. In \S\ref{res:evol} we found that different GMCs were effected by their environment in different ways, i.e. some experience mergers and others are sheared apart, etc.  Indeed recent observations of NGC 2276 have found that galactic-scale tidal forces and ram pressure has lead to large variations in molecular content of the galaxy, resulting in some regions with variations in the depletion time scale (the ratio of the molecular gas mass to the star formation rate) as large as several orders of magnitude when measured on scales of $\sim450\pc$ \citep{Tomicic:2018aa}. It is therefore likely that measurements of $\eff$ on cloud scales in such a galaxy also find large variations. We therefore argue that the initial, intrinsic properties of GMCs ($M_{\rm birth}$ and $R_{\rm birth}$) are not sufficient to set the initial value and evolution of $\eff$: other factors such as the galactic environment (e.g shear) must be taken into account.  It is the combination of  a wide range of possible cloud properties and the environment in which GMCs live that naturally give rise to the observed and simulated spread in $\eff$.

Additionally, we note that Fig.~\ref{fig:eff2Dhisttracks} and \ref{fig:masscuts} show that in our simulation $\eff$ does not follow a smooth, systematic evolution, instead it is able to both increase \emph{and} decrease as clouds age. This is a direct contradiction of the prediction made by the L16 model, which predicts a continually increase in $\efflee$ with $\tage$. The evolution seen in $\eff$ for our simulated clouds, as discussed above, result from combination of time dependant factors (e.g. mergers, gas accretion, feedback, shear and galactic tides) that are difficult to model as constant parameter in any given model and thus further evidence that the environment of clouds need to be accounted for when exploring their evolution. 

Finally, we note that the analytical model we have adopted from L16 is not the only model. For example, the model present in \cite{Vazquez:2018aa}  predicts different star formation histories (and hence instantaneous $\eff$) for clouds of different mass. Their results and conclusions support  this work and our conclusion that it is the evolution history of a cloud needs to be know to able to determine its $\eff$ at any given age. Therefore any model (numerical or analytical) must be able to capture the full range of physical processes that occur within a GMC and its interactions with its environment. 

\subsection{Limitations of the Simulations}
\label{dis:limits}

The instantaneous \HH fraction calculated at run-time to determine the star formation rate of a computational-cell (see \S\ref{sect:meth:sims}) is not stored or advected through the simulation. As a result we have to determine the molecular content within the simulation in post processing. For simplicity we chose to adapt a density threshold of $\rho_{\rm mol}=100\cc$, with all gas above this value considered to be molecular. This limits the maximum value of $\tff$, which in turn acts as a limiting factor in determining $\eff$ from the simulations. If larger values of $\tff$ could be reached smaller values of $\eff$ maybe detected. A simple solution would be to rerun the simulations but including a treatment of the chemistry and thus allowing the molecular fraction of the gas to be self-consistently determined by the simulation, which could then be used to identify GMCs. However given that the current simulation is able to not only reproduce the median value of $\eff$ but also the size and distribution of the spread in values (and does so using a universal efficiency on the scale of computational cells, i.e. $\effsf$) we leave re-simulation for future work.

Isolated GMC simulations are able to completely resolve the internal structure of the GMC but at the cost of the galactic environment \citep[e.g. see][]{Padoan:2016aa,Grudic:2018aa}. The simulations used throughout this work have such an environment but they are limited in spatial resolution (i.e. $\Delta x\sim4.6\pc$, see \S\ref{sect:meth:sims}). This resolution results in GMCs being made up of several computational cells and thus stars form and inject feedback into specific regions with the clouds. This allows one generation of stars to alter the gas structure within a GMC and even remove gas, thus determining where the next generation of stars form and how a GMC evolves. Despite the limited resolution, our simulation is able to accurately reproduce the galaxy wide gas probability distribution function (PDF), the range and distribution of cloud properties, including $\eff$ (see G17 and G18). We therefore argue that accurately fully resolving the internal structure of GMCs is not as important as accurately reproducing the large ($\geq100\pc$) scale galactic environment.

\section{Conclusions}
\label{sect:con}
In this work we explore the efficiency of Giant Molecular Clouds (GMCs) at forming stars in hydrodynamical simulations of Milky Way-like galaxies. The primary goal of this work is to explain the observed spread in the star formation efficiency per free fall time ($\eff$). To this end we calculate $\eff$ for each GMC found within two simulations, one with stellar feedback and one without. Using a tracking algorithm we follow the evolution of $\eff$ (and other properties) of each cloud throughout its lifetime. Our key results are:
\begin{enumerate}
 	\item Galactic disc simulations where star formation is determined by a Schmidt star formation law (applied to molecular gas) are able to produce the observed spread in the measured values of $\eff$ for GMCs. A large spread in values is found independently of the presence of stellar feedback, however the inclusion of feedback in the simulation prevents highly inefficient ($\eff<10^{-4}$) massive ($M_{\rm tot}>10^{7}\Msol$) clouds from forming. Stellar feedback is not the main source of the dispersion in $\eff$. 
	\item No single GMC property determines the $\eff$ of a cloud. Comparing seven key properties (gas mass, free-fall time, galactic radius, velocity dispersion, radius, density and virial parameter) of GMCs with $\eff$ shows no significant correlation. Instead we find that a cloud with a given value in any of the above properties is able to have a wide range of values in $\eff$. 
	\item Each GMC evolves in a unique way, determined by both its initial properties and its environment. It is therefore not possible to describe the evolution of all clouds by a single analytical model neglecting environmental effects. Furthermore, the evolution of a particular property for a given cloud is not smooth or uniform: a cloud is able to explore a wide range of values during its lifetime. This leads to a natural spread in properties and in particular the value of $\eff$.
	\item The evolution of $\eff$ throughout a GMC's lifetime does not follow a systematic increase, contrary to predictions of simple analytical models. Instead the measured value of $\eff$ for a cloud is driven by a number of time dependant factors, including stellar feedback and galactic environment,  which can cause both increases or decreases as the clouds ages. This allows for a variety of different star formation histories. 
\end{enumerate}

In future work we will explore the processes that drive cloud evolution and how this leads to diversity in cloud properties. 

\section*{acknowledgments}

We thank the anonymous referee for their valuable and insightful comments.
KG acknowledge support from the Science and Technology Facilities Council (grant ST/N002717/1), as part of the UK E-ELT Programme at the University of Oxford. 
KG also thanks and acknowledges support from New College, University of Oxford via the Balzan Fellowship.  
OA acknowledges support from the Swedish Research Council (grant 2014- 5791).  
OA and FR acknowledge support from the Knut and Alice Wallenberg Foundation. 
The research of JD and AS is supported by Adrian Beecroft and the STFC. 
This work used the DiRAC Complexity system, operated by the University of Leicester IT Services, which forms part of the STFC DiRAC HPC Facility (www.dirac.ac.uk). This equipment is funded by BIS National E-Infrastructure capital grant ST/K000373/1 and  STFC DiRAC Operations grant ST/K0003259/1. DiRAC is part of the National E-Infrastructure.

\bibliographystyle{mn3e}
\bibliography{ref}

\end{document}